# Performance of hybrid entanglement photon pair source for quantum key distribution

M. Fujiwara, M. Toyoshima and M. Sasaki

National Institute of Information and Communications Technology

4-2-1 Nukui-Kita, Koganei, Tokyo 184-8795, Japan

K. Yoshino, Y. Nambu, and A. Tomita

NEC Corporation, 34 Miyukigaoka, Tsukuba, Ibaraki 305-8501, Japan

Abstract:

We report on the experimental demonstration of a source of hybrid entanglement pairs between two different degrees of freedom, a 1550 nm time-bin qubit and an 810 nm polarization qubit. The polarization qubit at 810 nm is transformed by an asymmetric Mach-Zehnder interferometer consisted of a Glan laser prism and a polarization-maintaining fiber. We obtained visibilities of 95.8% and 88% with tolerance ±0.2 and ±1% along Z-Z and X-X axes on the Poincare sphere, respectively with a coincidence count rate of more than 800 c/s after entanglement format transformation. These values are well above the threshold of 70.7% needed to violate a Bell inequality and allow distilling a secure key in the quantum key distribution.

Main text:

Quantum entanglement is a fundamental resource for quantum communication. Its utilization, especially, in quantum key distribution (QKD) (1, 2) system is the most promising example developed out of quantum information. In the QKD system based on the sharing of entangled photons, only the detection units needed to be in possession of legitimate communication partners, while the entangled photons source needed not to be in a trusted hand. And in such an entanglement-based QKD system, we can avoid the need for fast quantum number generators that are indispensable for implementing the BB84 protocol. The entanglement-based QKD protocol also has potential for use in long distance communication in the presence of realistic experimental imperfections (i.e., dark counts of single photon detectors) compare to with weak coherent pulses. Moreover, entanglement swapping can naturally be used to extend to long distant and multi-party connections.

Two encoding schemes, time-bin encoding and polarization encoding, are often used to represent qubits. Specifically, time-bin encoding well suits for transmission in optical fibers at telecom wavelengths(3) because of its robustness against polarization mode disturbance. Indeed, long distance distribution of time-bin entangled qubits through an optical fiber has been demonstrated.(4, 5)

On the other hand, the transmission and measurement of polarization of the photon is straightforward to implement because of ease in encoding and decoding. So far there are many experimental implementations of entanglement distribution over free-space links at near infrared regions.(6-8) Such techniques will be of advantage in constructing flexible QKD networks. Furthermore recent experiments show that single-photon polarization qubits can be stored in two atomic ensembles. These techniques also have potential for realizing quantum repeaters.(9-11) The interface between the time-bin encoded photons and the polarization encoded ones will be profitable to build up a flexible QKD network combining fiber and free-space links. Such an interface that can be implemented with a hybrid entanglement photon pair source that generates a time-bin encoded 1550 nm photon and a polarization encoded 810 nm photon is highly desirable in quantum communication.

In a previous study, a hybrid entanglement was proposed and demonstrated. However, in the system, used in that study, both the receivers need polarization control.(12) We report on transformation from time-bin qubits to polarization qubits, and show the result of entanglement sharing between a receiver with standard polarization optics and one with a polarization independent planar lightwave circuit (PLC).

The hybrid entanglement source is constructed of a non-degenerate photon pair source and a format transformer from time encoding to polarization encoding. The non-degenerate down-conversion entangled photon pair source is shown in Fig. 1. Detailed information regarding its performance has been reported in Ref. 13. A 532 nm contiguous wave (CW) laser adjusted at 160 μW is incident on a 30 mm long periodically poled lithium niobate (PPLN) crystal that is quasi-phase-matched to create co-polarized entangled photons at 810 and 1550 nm. For this PPLN, the mode diameters of the 532 nm pump beam, 810 nm and 1550 nm photons are optimized to be 84 μm, 84 μm and 108 μm, respectively, to couple with single mode fibers.(14)

The 810 nm photons and the 1550 nm photons are separated using a dichroic mirror (DM). We use long-wavelength pass filters to reduce stray photons. After down conversion, a 1550 nm photon is incident a decoder set at Bob side. The decoder is a two-input and four-output silica-based PLC on a silicon substrate (15), featured by an asymmetric Mach-Zehnder interferometer (AMZI) with a 2.5 ns time delay. 50 cm long spiral pattern was fabricated to make the 2.5 ns time delay. This device was manufactured by NTT Electronics Corp. The output photons of the decoder are projected onto vectors $|0\rangle$, $|0\rangle-|1\rangle$, $|0\rangle+|1\rangle$, and $|1\rangle$ of the Poincare sphere (shown in the inset of Fig. 2), where $|1\rangle$ represents a photon in the first time-bin (having passed through the short arm), and $|0\rangle$ represents a photon in the second time-bin (having passed through the long arm). $|0\rangle$ and $|1\rangle$ are mapped on the polar state (z axis), and $|0\rangle-|1\rangle$ and $|0\rangle+|1\rangle$ are on the equatorial state (x axis). The AMZI is polarization insensitive by adjusting birefringence with precise temperature control.(15,16) This decoder has been used for a 100 km QKD test through a field installed fiber, and yielded stable interference over six hours without feedback control.(3) 1550 nm photons are detected by InGaAs based avalanche photo-diodes (APDs) with 10% detection efficiencies in the gated Geiger mode. The triggers for detection are given by Si based APDs (Perkin Elmer single photon counting modules: SPCMs) through a delay generator from Alice. The detection efficiency of the SPCM for 810 nm photons was about 55%. The entanglement based QKD system combining free space and fiber network can be realized as shown in Fig. 2. Hybridization of polarization information on the time-bin entanglement is described below.

Hybrid entanglement qubit is realized by superpositioning a polarization entanglement on a time-bin entanglement qubit. The 810 nm photon sent to Alice is input a format-transformer. Figure 3(a) and 3(b) show the schematic view and the equivalent optical circuit of the format transformer that consists of a polarizer set at 45° and a polarization sensitive AMZI constructed of a Glan laser prism and a polarization-maintaining (PM) fiber. In this AMZI, reflected photons from an escape window of the Glan laser prism are guided to the opposite side of the prism through a PM fiber with 2.5 ns time delay. The additional loss of the format transformer is 1.5dB. The format-transformer alters the entanglement state,

$$|\phi\rangle = \frac{1}{\sqrt{2}}[|0\rangle_A|0\rangle_B + \exp(i\theta(-\tau) - i\theta(0))|1\rangle_A|1\rangle_B] \tag{1}$$

to the one described by

$$|\phi\rangle = \frac{1}{\sqrt{2}}[|H\rangle_A|0\rangle_B + \exp(i\theta(-\tau) - i\theta(0))|V\rangle_A|1\rangle_B] \qquad (2)$$

where, $|H\rangle$ and $|V\rangle$ represent horizontal and vertical polarization states, respectively. The indices A and B stand for Alice and Bob. The relative phase $\theta(t)$ is defined with respect to a reference path length difference $\tau$ between the short and the long arms.

The polarization states of Alice's photons can be analyzed with a half wave plate, polarization beam splitters (PBSs) and SPCMs. The polar states (z axis), $|0\rangle$ or $|1\rangle$, can be easily distinguished by using a PBS because they are transformed to $|H\rangle$ or $|V\rangle$. On the other hand, the equatorial states (x axis) $|0\rangle$-$|1\rangle$ and $|0\rangle$+$|1\rangle$ are analyzed after through a half wave plate set at 22.5°. The entanglement state given in eq.2 changes as follows;

$$|\phi\rangle = \frac{1}{2}[|-\rangle_A(|0\rangle_B - \exp(i\theta(-\tau) - i\theta(0))|1\rangle_B) + |+\rangle_A(|0\rangle_B + \exp(i\theta(-\tau) - i\theta(0))|1\rangle_B)]$$
$$= \frac{1}{2}[|-\rangle_A(|0\rangle_B - |1\rangle_B) + |+\rangle_A(|0\rangle_B + |1\rangle_B)] \qquad (3)$$

where $|-\rangle_A = \frac{1}{\sqrt{2}}(|H\rangle_A - |V\rangle_A)$, and $|+\rangle_A = \frac{1}{\sqrt{2}}(|H\rangle_A + |V\rangle_A)$. Equation 3 clearly describes the entanglement in the equatorial state (x axis), and it can also be analyzed by using a PBS and an SPCM.

To demonstrate the presence of this entanglement, we measured the coincidence events of the detection of the 810 nm and 1550 nm photons in the appropriate time window (40 nsec). The triggers for the gate pulses were given by OR of the detection outputs of the SPCMs at Alice, and fed to Bob's APDs through a delay generator.

Figure 4(a) shows the coincidence counts of the polar state (z axis) as a function of delay time. The detector named Z0(Z1) stands for the detector of signal 0(1) on the z axis. Coincidence of Z0(Alice)-Z0(Bob) and Z1(Alice)-Z1(Bob) should be detected at the same delay time. Thus photons in the central peaks correspond to photon pairs created in the first (second) time-bin which pass through the short (long) arms at a format transformer and an AMZI for a 1550 nm photon. It can be observed that the maximum values of these pairs are obtained at almost the same delay time and the visibilities are 95.8±0.2%. The total coincidence count rate is 820 c/s.

The coincidence counts in the equatorial state (x axis) are shown in Fig. 4(b). AMZIs in a format transformer and for 1550 nm photon have to be correctly adjusted so that a high correlation can be obtained between Alice and Bob. In the present setup, the phases are adjusted by changing the PLC temperature at Bob's side. The visibilities of 88±1% are achieved. The coincidence count rate is 950 c/s.

Both these results violate the Bell's inequality of 70.7% for entanglement distribution, allowing the distillation of a secure key for QKD. To increase the coincidence count rate of this system, decreasing a time delay in the AMZI is the most effective way. However, fabrication of a PLC with short delay time keeping polarization independent characteristics becomes difficult because a short delay line makes polarization adjustability of the PLC small and consequently aligning the optimal temperature within the operation range of a controller becomes difficult. Therefore, sophisticated process technique should be developed to improve our system.

We have demonstrated a source of hybrid entanglement pairs. The entanglement is hybrid in the sense that a time-bin qubit at 1550 nm is entangled with a polarization qubit at 810 nm. We obtained visibilities of 95.8% and 88% with tolerance ±0.2 and ±1% along Z-Z and X-X axes on the Poincare sphere respectively, with a coincidence count rate of greater than 800 c/s. This is well above the threshold of 70.7% needed to violate a Bell inequality and also allows distilling a secure key in QKD. The PLC technology enabled us to realize a stable time-bin qubit decoder with a small footprint, which will be also useful to implement quantum circuits for photonic quantum computation.(17) It should be noted that the decoder at the free space receiver does not need to equip an interferometer; only the polarization analyzer is required. Thus our scheme simplifies the free-space receiver structure to be robust to the temperature changes and mechanical disturbances. The free-space receiver will be suitable for use as a mobile terminal in a space-fiber hybrid communication link. This research will also open the road for future demonstrations of more complicated quantum communication protocols such as entanglement swapping.

Figure captions

Fig. 1. The hybrid entanglement photon pair source. A second harmonic generation of YAG CW 160 μW laser (532 nm) is used as a pump light. The length of the PPLN is 30 mm. Input and output ends of the PPLN are AR coated. The 810 nm photons and the 1550 nm photons are separated using a dichroic mirror. Long-wavelength pass filters with cutoff wavelengths of 680 nm and 900 nm are used for 810 nm and 1550 nm photon respectively.

Fig. 2. Conceptual view of the QKD system setup with the hybrid entanglement photon pair source. Inset shows the correspondence between a time-bin and a polarization on the Poincare sphere.

Fig. 3. (a) Conceptual view and (b) equivalent optical circuit of the format transformer. The first (second) time-bin which passes through the short (long) arms of a format transformer is superimposed the polarization information as $|H\rangle$ ($|V\rangle$). Instead of a normal PBS, the Glan laser prism is selected to improve an extinction ratio.

Fig. 4. Coincidence counts of (a) polar (z axis) state as a function of delay time, and (b) equatorial (x axis) state as a function of the PLC operation temperature. Coincidence of Z0(Alice)-Z0(Bob) and Z1(Alice)-Z1(Bob) are detected in center peaks of (a) . And the visibilities are 95.8±0.2% are obtained. The visibilities of (b) are 88±1%, calculated from fringes.

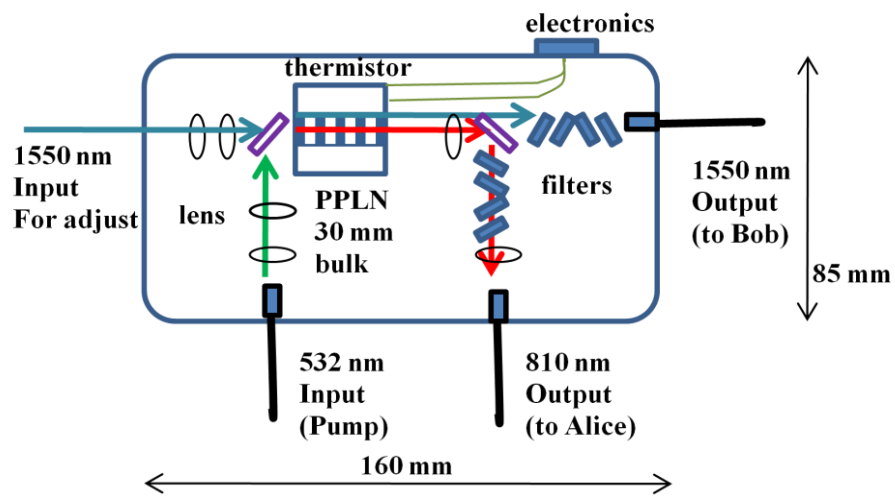

Fig. 1

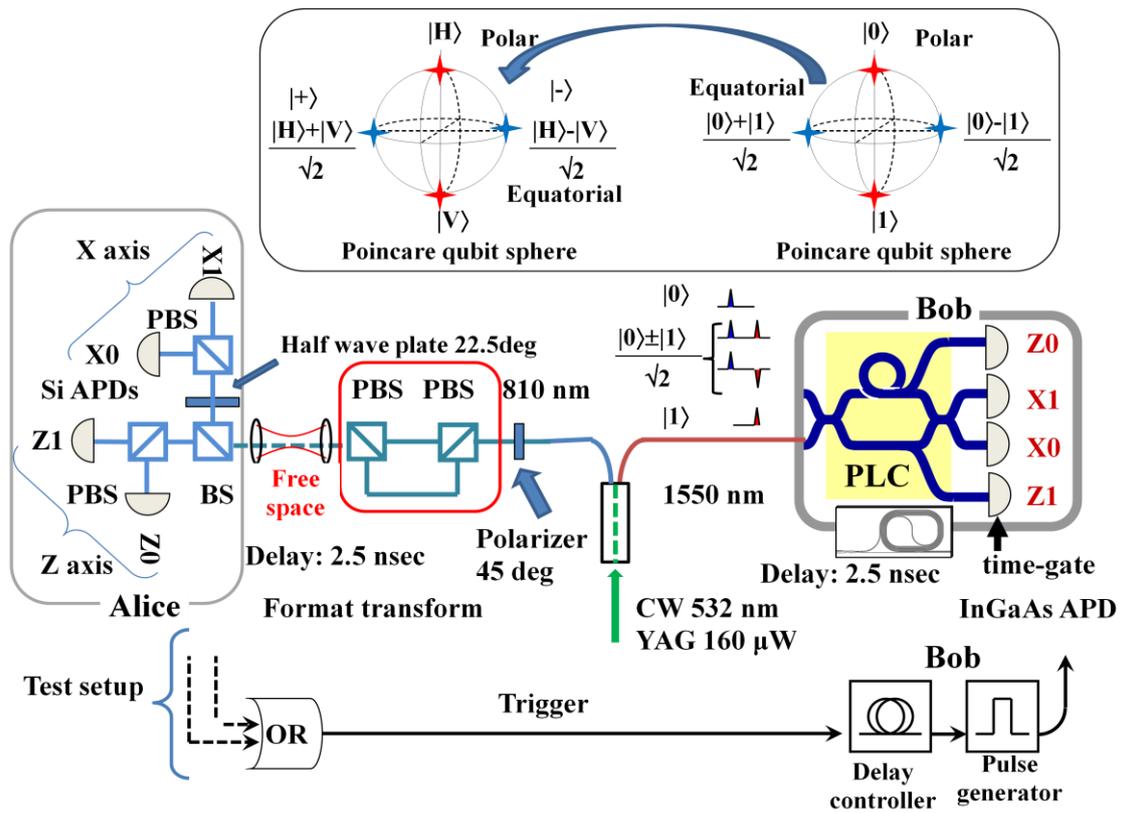

Fig.2

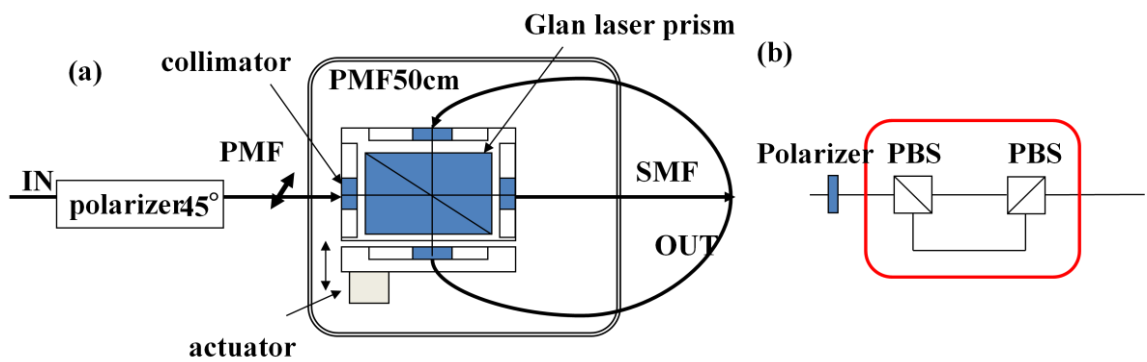

Fig. 3

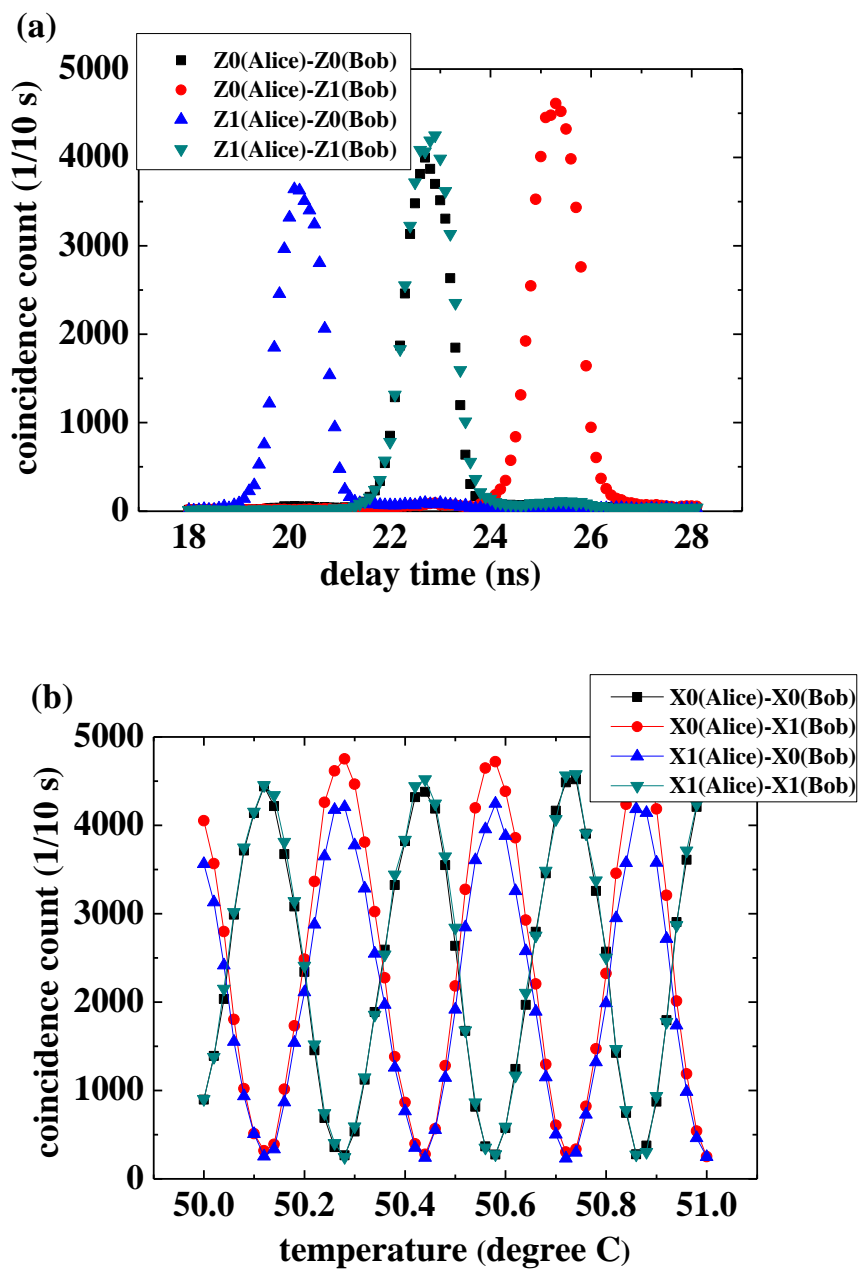